\documentclass[11pt]{amsart}

\usepackage{amscd}
\usepackage{amsmath}
\usepackage{graphicx}
\usepackage{amsfonts}
\usepackage{amssymb}
\begin{document}

\title[Optimal entanglement witnesses]
{A characterization of optimal entanglement witnesses}

\author{Xiaofei Qi}
\address[Xiaofei Qi]{
Department of Mathematics, Shanxi University , Taiyuan 030006, P. R.
of China;} \email{qixf1980@126.com}

\author{Jinchuan Hou}
\address{Department of
Mathematics\\
Taiyuan University of Technology\\
 Taiyuan 030024,
  P. R. of China}
\email{houjinchuan@tyut.edu.cn; jinchuanhou@yahoo.com.cn}


\thanks{This work is partially supported by  Research Fund for the Doctoral Program
of Higher Education of China (20101402110012),  A grant from
International Cooperation Program in Sciences and Technology  of
Shanxi (2011081039), Tianyuan Funds of China (11026161) and Natural
Science Foundation of China (11171249, 11101250).}

\begin{abstract}

In this paper, we present a characterization of optimal entanglement
witnesses in terms of positive maps and then provide a general
method of checking optimality of entanglement witnesses. Applying
it, we obtain  new indecomposable optimal witnesses which have no
spanning property. These  also provide new examples which support a
recent conjecture saying that the so-called structural physical
approximations to optimal positive maps (optimal entanglement
witnesses) give entanglement breaking maps (separable states).

\end{abstract}
\maketitle

\section{Introduction}

Let $H$  be a separable complex Hilbert space. Recall that a quantum
state on $H$ is a density operator $\rho\in{\mathcal B}(H)$ (the von
Neumann algebra of all bounded linear operators) which is positive
and has trace 1. Denote by ${\mathcal S}(H)$ the set of all states
on $H$. If $H$ and $K$ are finite dimensional, a state in the
bipartite composition system  $\rho\in{\mathcal S}(H\otimes K)$ is
said to be separable if $\rho$ can be written as $\rho=\sum_{i=1}^k
p_i \rho_i\otimes \sigma _i,$ where $\rho_i$ and $\sigma_i$ are
states on $H$ and $K$ respectively, and $p_i$ are positive numbers
with $\sum _{i=1}^kp_i=1$.  For the case that at least one of $H$
and $K$ is of infinite dimension, a state $\rho$ acting on $H\otimes
K$ is called separable if it can be approximated in the trace norm
by the states of the form $\sigma=\sum_{i=1}^n p_i \rho_i\otimes
\sigma _i,$ where $\rho_i$ and $\sigma_i$ are states on $H$ and $K$
respectively, and $p_i$ are positive numbers with
$\sum_{i=1}^np_i=1$. Otherwise, $\rho$ is said to be inseparable or
entangled (ref. \cite{BZ, NC}).

Entanglement is a basic physical resource to realize various quantum
information  and quantum communication tasks such as quantum
cryptography, teleportation, dense coding and key distribution
\cite{NC}. It is very important but also difficult to determine
whether or not a state in a composite system is separable. A most
general approach to characterize quantum entanglement is based on
the notion of entanglement witnesses (see \cite{Hor}). A Hermitian
(i.e., self-adjoint) operator $W$ acting on $H\otimes K$ is  an
entanglement witness (briefly, EW) if $W$ is not positive and ${\rm
Tr}(W\sigma)\geq 0$ holds for all separable states $\sigma$. Thus,
if $W$ is an EW, then there exists an entangled state $\rho$ such
that ${\rm Tr}(W\rho)<0$ (that is, the entanglement of $\rho$ can be
detected by $W$). It was shown that, a state is entangled if and
only if it is detected by some entanglement witness \cite{Hor}.
However, constructing entanglement witnesses is a hard task. There
was a considerable effort in constructing and analyzing the
structure of entanglement witnesses for both finite and infinite
dimensional systems \cite{B,TG,CK,JB,HQ}. However, complete
characterization and classification of EWs is far from satisfactory.

Due to the Choi-Jamio{\l}kowski isomorphism \cite{C,J}, a
self-adjoint operator $W\in{\mathcal B}(H\otimes K)$ with $\dim
H\otimes K<\infty$ is an EW if and only if there exists a positive
linear map which is not completely positive (NCP) $\Phi:{\mathcal
B}(H)\rightarrow{\mathcal B}(K)$ and a maximally entangled state
$P^+\in{\mathcal B}(H\otimes H)$ such that $W=W_\Phi=(I_n\otimes
\Phi)P^+$. Recall that a maximally entangled state is a pure state
$P^+=|\psi^+\rangle\langle\psi^+|$ with
$|\psi^+\rangle=\frac{1}{\sqrt{n}}(|11\rangle+|22\rangle+\cdots
|nn\rangle) $, where $n=\dim H$, $\{|i\rangle\}_{i=1}^n$ is an
orthonormal basis of $H$. Thus, up to a multiple by positive scalar,
$W_\Phi$ can be written as the matrix $W_\Phi=(\Phi(E_{ij}))$, where
$E_{ij}=|i\rangle\langle j|$. For a positive linear map
$\Phi:{\mathcal B}(H)\rightarrow{\mathcal B}(K)$, we always denote
$W_\Phi$  the Choi-Jamio{\l}kowski matrix of $\Phi$ with respect to
a given basis of $H$, that is $W_\Phi=(\Phi(E_{ij}))$, and we say
that  $W_\Phi$ is the witness associated to the positive map $\Phi$.
Conversely, for an entanglement witness $W$, we denote $\Phi_W$ for
the associated positive map so that $W=W_{\Phi_W}$.

For any entanglement witness $W$,  let
$$\mathcal{D}_W=\{\rho:\rho\in{\mathcal S}(H\otimes K), {\rm Tr}(W\rho)<0\},$$
that is, ${\mathcal D}_W$ is the set of all entangled states that
detected by $W$. For entanglement witnesses $W_1,W_2$, we say that
$W_1$ is finer than $W_2$ if $\mathcal{D}_{W_2}\subset
\mathcal{D}_{W_1}$, denoted by $W_2\prec W_1$. While, an
entanglement witness $W$ is optimal  if there exists no other
witness finer than it. Obviously, a state $\rho$ is entangled if and
only if there is some optimal EW such that ${\rm Tr}(W\rho)<0$. In
\cite{LK}, Lewenstein, Kraus, Cirac and Horodecki proved that: (1)
$W$ is an optimal entanglement witness if and only if $W-Q$ is no
longer an entanglement witness for arbitrary positive operator $Q$;
(2) $W$ is optimal if ${\mathcal P}_W= \{|e,f\rangle\in H\otimes K:
\langle e, f|W|e,f\rangle=0\}$ spans the whole $H\otimes K$ (in this
case, we say that $W$ has spanning property). For the infinite
dimensional version of these results, see \cite{HG}. To the best of
the author's knowledge, the above criterion (2) is the only method
we have known by now that is practical  of checking optimality of
witnesses. In fact, almost all known optimal EWs are checked by
using of the  criterion (2) (Ref. \cite{CW,HK} and the references
therein). However, the criterion is only a sufficient condition.
There are known optimal witnesses that have no spanning property.
For example, consider the Choi map $\phi$ from $M_3$ into $M_3$
defined by
$$ \left(\begin{array}{ccc} a_{11}&a_{12}&a_{13}\\
a_{21}&a_{22}&a_{23}\\ a_{31}&a_{32}&a_{33}\end{array}\right)\mapsto\left(\begin{array}{ccc} a_{11}+a_{33}&-a_{12}&-a_{13}\\
-a_{21}&a_{22}+a_{11}&-a_{23}\\
-a_{31}&-a_{32}&a_{33}+a_{22}\end{array}\right). \eqno(0.1)
$$ It is well known that the associated entanglement witness $W_{\phi}$ is optimal (by proving that $\phi$ is an extremal point of the convex set
of all completely positive linear maps on $M_3$) and span$({\mathcal
P}_{W_{\phi}})\not={\mathbb C}^3\otimes {\mathbb C}^3$. Thus a
natural question rises: Are there any other practical methods to
detect the optimality of  entanglement witnesses?

The purpose of this paper is to give a necessary and sufficient
condition for an EW to be  optimal  in terms of positive maps. Based
on this result, we  give a general approach of how to check that an
EW is optimal or not. This approach is practical. Applying it we
show that the entanglement witnesses arising from the positive maps
in \cite{QH} are indecomposable optimal witnesses. Moreover, these
optimal EWs give new examples supporting a recent conjecture posed
in \cite{KABL} saying that the so-called structural physical
approximations (SPA) to optimal positive maps (optimal EWs) give
entanglement breaking  maps (separable states).

Recall that an entanglement witness $W$ is called decomposable if
$W =Q_1+Q_2^{\Gamma}$ for some operators $Q_1,Q_2\geq0$, where
$Q_2^{\Gamma}$ stands for any one of $Q_2^{T_1}$ and $Q_2^{T_2}$,
 the partial transpose of $Q_2$ with respect to the
subsystem $H$ and $K$, respectively; a positive map $\Delta$ is said
to be decomposable if it is the sum of a completely positive map
$\Delta_1$ and the composition of a completely positive map
$\Delta_2$ and the transpose $\bf T$, i.e.,
$\Delta=\Delta_1+\Delta_2\circ{\bf T}$. Hence $W_\Phi$ is
decomposable if and only if $\Phi$ is decomposable. A completely
positive map $\Lambda$ is called entanglement breaking (EB) if its
partial action $\Lambda \otimes I$ sends every state to a separable
state.

Throughout this paper, $H$ and $K$ are  complex Hilbert spaces, and
$\langle\cdot|\cdot\rangle$ stands for the inner product in both of
them. ${\mathcal B}(H,K)$ (${\mathcal B}(H)$ when $K=H$) is the
Banach space of all (bounded linear) operators from $H$ into $K$.
$A\in{\mathcal B}(H)$ is self-adjoint if $A=A^\dagger$ ($A^\dagger$
stands for the adjoint operator of $A$); and $A$ is positive,
denoted by $A\geq 0$, if
 $\langle \psi | A|\psi\rangle\geq 0$ for all $|\psi\rangle\in H$.
For any positive integer $n$, $H^{(n)}$ denotes the direct sum of
$n$ copies of $H$.   A linear map $\Phi$ from ${\mathcal B}(H)$ into
${\mathcal B}(K)$   is positive if $A\in{\mathcal B}(H)$ is positive
implies that $\Phi(A)$ is positive; $\Phi$ is $k$-positive if
$\Phi_k=\Phi\otimes I_k: {\mathcal B}(H)\otimes M_k\rightarrow
{\mathcal B}(K)\otimes M_k$ defined by $\Phi_k((A_{ij})_{k\times
k})=(\Phi(A_{ij}))_{k\times k}$ is positive; $\Phi$ is completely
positive (CP) if $\Phi_k$ is positive for all positive integers $k$.
By Choi's well-known result, if $H$ and $K$ are finite dimensional,
then $\Phi$ is completely positive if and only if
$W_\Phi=(\Phi(E_{ij}))$ is a positive (semi-definite) matrix. A
linear map $\Phi: {\mathcal B}(H)\rightarrow{\mathcal B}(K)$ is
called an elementary operator if there are two finite sequences
$\{A_i\}^n_{i=1}\subset{\mathcal B}(H,K)$ and
$\{B_i\}^n_{i=1}\subset{\mathcal B}(K,H)$ such that
$\Phi(X)=\sum_{i=1}^n A_iXB_i$ for all $X\in{\mathcal B}(H)$. Note
that every linear map from ${\mathcal B}(H)$ into ${\mathcal B}(K)$
is an elementary operator if $H$ and $K$ are finite dimensional.

\section{A characterization of the optimality of entanglement witnesses}

In this section we first give a characterization of optimality of
EWs in terms of positive elementary operators. Then, by using of the
result, we develop a general approach how to check the optimality of
entanglement witnesses.

Before stating the main results in this section, let us recall
some notions and give a lemma from \cite{H4}.

Let $l$, $k\in\mathbb{N}$ (the set of all natural numbers),  and let
$A_{1},\cdots, A_{k}$, and $C_{1},\cdots, C_{l}\in {\mathcal B}(H$,
$K$). If, for each $|\psi\rangle\in H$, there exists an $l\times k$
complex matrix $(\alpha _{ij}(|\psi\rangle))$ (depending on
$|\psi\rangle$) such that
$$
C_{i}|\psi\rangle=\sum _{j=1}^{k}\alpha
_{ij}(|\psi\rangle)A_{j}|\psi\rangle,\qquad i=1,2,\cdots ,l,
$$
we say that $(C_{1},\cdots ,C_{l})$ is a locally linear combination
of $(A_{1},\cdots ,A_{k})$, $(\alpha_{ij}(|\psi\rangle))$ is called
a {\it local coefficient matrix} at $|\psi\rangle$.  Furthermore, if
a local coefficient matrix $(\alpha_{ij}(|\psi\rangle))$
  can be chosen for every $|\psi\rangle\in H$ so that
its operator norm $\|(\alpha _{ij}(|\psi\rangle))\|\leq 1$, we say
that $(C_{1},\cdots ,C_{l})$ is a {\it contractive locally linear
combination} of $(A_{1},\cdots ,A_{k})$; if there is a matrix
$(\alpha_{ij})$ such that $C_{i}=\sum _{j=1}^{k}\alpha _{ij}A_{j}$
for all $i$, we say that $(C_{1},\cdots ,C_{l})$ is a {\it linear
combination} of $(A_{1},\cdots ,A_{k})$ with coefficient matrix
$(\alpha _{ij})$. Sometimes we also write $\{A_i\}_{i=1}^k$ for
$(A_{1},\cdots ,A_{k})$.

The following characterization of  positive elementary operators
was obtained in \cite{H4}, also, see \cite{H}.

\textbf{Lemma 2.1.}  {\it Let $H$ and $K$ be complex Hilbert
spaces of any dimension, $\Phi: {\mathcal
B}(H)\rightarrow{\mathcal B}(K)$ be a linear map defined by
$\Phi(X) =\sum _{i=1}^{k}C_{i}XC_{i}^{\dagger}-\sum
_{j=1}^{l}D_{j}XD_{j}^{\dagger}$ for all $X$. Then $\Phi $ is
positive if and only if $(D_{1},\cdots ,D_{l})$ is a contractive
locally linear combination of $(C_{1},\cdots ,C_{k})$.
Furthermore, $\Phi$  is completely positive if and only if
$(D_{1},\cdots ,D_{l})$ is a linear combination of $(C_{1},\cdots
,C_{k})$ with a contractive coefficient matrix, and in turn, if
and only if there exist $E_1, E_2, \ldots , E_r$  in ${\rm
span}\{C_{1},\cdots ,C_{k}\}$  such that $ \Phi=\sum_{i=1}^r
E_i(\cdot )E_i^\dagger.$}

Since every linear map between matrix algebras is an elementary
operator, and every hermitian preserving linear map is of the form
 $\Phi(\cdot) =\sum _{i=1}^{k}C_{i}(\cdots)C_{i}^{\dagger}-\sum
_{j=1}^{l}D_{j}(\cdot)D_{j}^{\dagger}$,  Lemma 2.1 gives a
characterization of positive maps from ${\mathcal B}(H)$ into
${\mathcal B}(K)$ in the case that both $H$ and $K$ are finite
dimensional.

The following is the main result of this section.

{\bf Theorem 2.2.} {\it Let $H$ and $K$ be finite dimensional
complex Hilbert spaces. Let $\Phi:{\mathcal B}(H)\rightarrow
{\mathcal B}(K)$ be a positive linear map. Then $W_\Phi$ is an
optimal entanglement witness if and only if, for any $C\in {\mathcal
B}(H,K)$, the map $X\mapsto \Phi(X)-CXC^\dag$ is not a positive
map.}

{\bf Proof.} Fix an orthonormal basis of $H$ and $K$, respectively.
Assume that $W_\Phi$ is optimal, by \cite{LK}, $W_\Phi-D$ is not an
entanglement witness for any nonzero $D\geq 0$. Take any $C\in
{\mathcal B}(H,K)$ and consider the map $A\mapsto \Phi(A)-CAC^\dag$.
Since the map $A\mapsto CAC^\dag$ is completely positive, the
corresponding Choi-Jamio{\l}kowski matrix $W_C$ is positive. The
optimality of $W_\Phi$ implies that $W_\Phi-W_C$ is not an EW, and
so the map $A\mapsto \Phi(A)-CAC^\dag$ is not positive.

On the other hand,  if the map $X\mapsto\Phi(X)-CXC^\dag$ is not a
positive map for any $C\in{\mathcal B}(H,K)$, we will show that
$W_\Phi$ is optimal. By the Choi-Jamio{\l}kowski isomorphism, any
positive operator $D$ corresponds to a completely positive linear
map $\Phi_D$. Thus there exist non-zero operators $E_1,\ldots, E_k$
such that $\Phi_D(X)=\sum_{i=1}^kE_iXE_i^\dag$ for all $X$. By the
assumption, $X\mapsto\Phi(X)-E_1XE_1^\dag$ is not a positive map,
and hence the map $X\mapsto\Phi(X)-\sum_{i=1}^kE_iXE_i^\dag$ is not
positive either. So $W_\Phi-D$ is not an EW for any positive
operator $D\in{\mathcal B}(H,K)$, which implies that $W_\Phi$ is
optimal. \hfill$\Box$

 {\bf Corollary 2.3.} {\it Let  $\Phi:{\mathcal B}(H)\rightarrow
{\mathcal B}(K)$ be a positive linear map defined by
$\Phi(X)=\sum_{i=1}^kA_iXA_i^\dag-\sum_{j=1}^lB_jXB_j^\dag$ for all
$X$. Then  $W_\Phi$ is optimal if and only if, for any operator
$C\in{\mathcal B}(H,K)$ which is a contractive locally linear
combination of $\{A_i\}_{i=1}^k$, the map $\Psi$ defined by
$\Psi(X)=\Phi(X)-CXC^\dag$ is not positive.}

{\bf Proof.} The ``only if" part is clear by Theorem 2.2. For the
``if" part, assume that $C\in{\mathcal B}(H,K)$  is not a
contractive locally linear combination of $\{A_i\}_{i=1}^k$. By
Lemma 2.1, the map $\Psi$ defined by $\Psi(X)=\Phi(X)-CXC^\dag$ is
not positive. So, together with the hypotheses, we see that the map
$X\mapsto\Phi(X)-CXC^\dag$ is not a positive map for any $C$. Then,
by Theorem 2.2, $W_\Phi$ is an optimal EW. \hfill$\Box$

By Corollary 2.3, we provide a method of checking the optimality of
an entanglement witness.

 {\bf A general approach of checking optimality.}
Assume that $\dim H=n$ and $\dim K=m$. Identify  $H$ and $K$ as
${\mathbb C}^n$ and ${\mathbb C}^m$, respectively. If $W$ is an EW
of the system $H\otimes K$, then there exists some NCP positive
linear map $\Phi:M_n\rightarrow M_m$ such that
$W=W_\Phi=(\Phi(E_{ij}))$. By Lemma 2.1, $\Phi$ has the form
$\Phi(X) =\sum _{i=1}^{k}C_{i}XC_{i}^{\dagger}-\sum
_{j=1}^{l}D_{j}XD_{j}^{\dagger}$ for all $X\in M_n$, where
$(D_{1},\cdots ,D_{l})$ is a contractive locally linear combination
of $(C_{1},\cdots ,C_{k})$. To check the optimality of $W$, let
$C\in {\mathcal B}({\mathbb C}^n, {\mathbb C}^m)$ be any operator
such that $C$ is a contractive locally linear combination of
$(C_{1},\cdots ,C_{k})$. If $X\mapsto\Phi(X)-CXC^\dag$ is positive,
then, by Lemma 2.1, for any $|x\rangle\in {\mathbb C}^n$,  there
exist scalars $\{ \alpha_{ij}(|x\rangle)\}_{i=1;j=1}^{k;l}$ and
$\{\gamma_i(|x\rangle)\}_{i=1}^k$ such that
$D_j|x\rangle=\sum_{i=1}^k\alpha_{ij}(|x\rangle)A_i|x\rangle$,
$C|x\rangle=\sum_{i=1}^k\gamma_i(|x\rangle)A_i|x\rangle$, and the
matrix $$F(|x\rangle)=\left(\begin{array}{cccc}
 \alpha_{11}(|x\rangle) & \alpha_{12}(|x\rangle)& \cdots &\alpha_{1k}(|x\rangle) \\
 \alpha_{21}(|x\rangle) & \alpha_{22}(|x\rangle)& \cdots &\alpha_{2k}(|x\rangle) \\
\vdots& \vdots& \ddots &\vdots \\
\alpha_{l1}(|x\rangle) & \alpha_{l2}(|x\rangle)& \cdots &\alpha_{lk}(|x\rangle) \\
\gamma_1(|x\rangle) & \gamma_2(|x\rangle)& \cdots
&\gamma_k(|x\rangle)
\end{array}\right)$$ is contractive. Thus, $W=W_\Phi$ is optimal if
and only if, for any operator $C$, there exists some vector
$|x\rangle$ such that $\|F(|x\rangle)F(|x\rangle)^\dag\|>1$ for all
possible choice of coefficient matrices $F(|x\rangle)$.

This approach is very useful especially for  those entanglement
witnesses $W$ with span$({\mathcal P}_W)\not=H\otimes K$. In the
next section, we will use this method to show that
$W_{\Phi^{(n,k)}}$  ($n\geq 3$ and $k=1,2,\ldots, n-1$)  are
indecomposable optimal entanglement witnesses if
$k\not=\frac{n}{2}$, where $\Phi^{(n,k)}$s are NCP positive maps
constructed in \cite{QH}.

\section{Optimality of some indecomposable entanglement witnesses}

The following  kind of  NCP positive linear maps
$\Phi^{(n,k)}:M_n({\mathbb C})\rightarrow M_n({\mathbb C})$ are
constructed in \cite{QH},
 $$\Phi^{(n,k)}
(A)=(n-1)\sum_{i=1}^nE_{ii}AE_{ii}+\sum_{i=1}^{n}E_{i,\pi^k(i)}AE_{\pi^k(i),i}-
A \eqno(3.1) $$ for every $A\in M_n$, where $n\geq 3$ and
$k=1,2,\cdots, n-1$, $E_{ij}$ are the matrix units as usual and
$\pi^1=\pi$ is a permutation of $\{1,2,\cdots ,n\}$ defined by
$\pi(i)=(i+1)\ {\rm mod} \ n$, $\pi^k(i)=(i+k)\ {\rm mod} \ n$
($k>1$), $i=1,2,\cdots, n$. That is, $\Phi^{(n,k)}$ maps $n\times n$
matrix $(a_{ij})$ to ${\rm
diag}((n-1)a_{11}+a_{k+1,k+1},(n-1)a_{22}+a_{k+2,k+2},\cdots,
(n-1)a_{nn}+a_{kk})-(a_{ij}).$ Moreover, it was shown in \cite{QH}
that $\Phi^{(n,k)}$ is indecomposable whenever either $n$ is odd or
$n$ is even but $k\neq \frac{n}{2}$. For the case $n=3$ and $k=2$,
one gets the Choi map $\phi=\Phi^{(3,2)}$ defined by Eq.(0.1).
 The purpose of this section is to show, by using of
the approach provided in the previous section, that all
$W_{\Phi^{(n,k)}}$s are indecomposable optimal entanglement
witnesses whenever $k\not=\frac{n}{2}$; while in the case $n$ is
even, $W_{\Phi^{(n,\frac{n}{2})}}$ is decomposable and not optimal.

The following  lemma  is obvious but useful to our purpose.

{\bf Lemma 3.1.} {\it Assume that $F=\left(\begin{array}{cc}
 1 & b\\
 \bar{b} &a\end{array}\right)\in M_2({\mathbb C})$ is positive semi-definite. If $b\not=0$, then
 $\|F\|>1$.}

{\bf Theorem 3.2.} {\it For $n\geq 3$, $k=1,2,\ldots, n-1$, let
$\Phi^{(n,k)}:M_n({\mathbb C})\rightarrow M_n({\mathbb C})$ be the
positive linear maps defined by Eq.(3.1). Then}

(1) {\it the entanglement witness $W_{\Phi^{(n,k)}}$ is
indecomposable and optimal whenever $k\not=\frac{n}{2}$;}

(2) {\it  the entanglement witness $W_{\Phi^{(n,\frac{n}{2})}}$ is
decomposable and not optimal, in this case $n\geq 4$ is an even
integer.}

{\bf Proof.}  We first prove the assertion (1). We give the
details of proof for the maps $\Phi=\Phi^{(n,1)}$. Other
$\Phi^{(n,k)}$s are dealt with similarly whenever
$k\not=\frac{n}{2}$ in the case that $n$ is even.

It is clear that $W_{\Phi}$ is indecomposable as $\Phi$ is
indecomposable by \cite{QH}. In the sequel we show that $W_\Phi$ is
also optimal by using of the  approach presented in Section 2.

Take any $C\in M_{n}$ and let
$$\begin{array}{rl}\Psi_C(A)=&\Phi(A)-CAC^\dag\\
=&(n-1)\sum_{i=1}^nE_{ii}AE_{ii}^\dagger
+\sum_{i=1}^{n}E_{i,i+1}AE_{i,i+1}^\dagger-A-CAC^\dag\end{array}$$for
all $A\in M_n$. By Corollary 2.3, we only need to consider the case
that  $C$ is a contractive locally linear combination of
$$\{ \sqrt{n-1}E_{11}, \sqrt{n-1}E_{22},\cdots, \sqrt{n-1}E_{nn},
E_{12}, E_{23},\cdots, E_{n1} \}.$$

Since $\Phi$ is positive, by Theorem 2.1, for any
$|x\rangle=(x_1,x_2,\ldots,x_n)^T$, there exist scalars
$\alpha_1,\alpha_2,\cdots, \alpha_n,\beta_1,\beta_2,\cdots,\beta_n$
(depending on $|x\rangle$) with
$\sum_{i=1}^n(|\alpha_i|^2+|\beta_i|^2)\leq 1$  such that
$$|x\rangle=\sum_{i=1}^n\alpha_i
(\sqrt{n-1}E_{ii})|x\rangle+\sum_{i=1}^n\beta_iE_{i,i+1}|x\rangle.\eqno(3.2)$$
Consider the case that $x_i\not=0$ for all $i$. By Eq.(3.2), we get
$x_i=\sqrt{n-1}\alpha_i x_i+\beta_i x_{i+1}$, and so
$$\beta_i
=(1-\sqrt{n-1}\alpha_i)\frac{x_i}{x_{i+1}}\ \ {\rm for\ \ all} \ \
i=1,2,\cdots,n.\eqno(3.3)$$ Write
$|x\rangle=(|x_1|e^{i\theta_1},|x_2|e^{i\theta_2},\cdots,|x_n|e^{i\theta_n})^T$
and let $r_i=|\frac{x_i}{x_{i+1}}|^2$ for all $i=1,2,\cdots,n$.
Define a function
$$\begin{array}{rl}
f(\alpha_1,\alpha_2,\cdots,\alpha_n)
=&\sum_{i=1}^n|\alpha_i|^2+\sum_{i=1}^n|\beta_i|^2\\
=&\sum_{i=1}^n|\alpha_i|^2+\sum_{i=1}^n|1
-\sqrt{n-1}\alpha_i|^2r_i.\end{array}$$ For every $j$, write
$\alpha_j=a_j+ib_j$ with $a_j$ and $b_j$ real. Then the above
equation reduces to
$$\begin{array}{rl}
f(\alpha_1,\alpha_2,\cdots,\alpha_n)
=&\sum_{i=1}^na_i^2+\sum_{i=1}^nb_i^2+\sum_{i=1}^n(n-1)a_i^2r_i\\&
+\sum_{i=1}^n(n-1)b_i^2r_i+\sum_{i=1}^nr_i-2\sqrt{n-1}\sum_{i=1}^na_ir_i,
\end{array}\eqno(3.4)$$
where $\prod_{i=1}^n r_i=1$ with $r_i>0$ for $i=1,2,\cdots,n$.

Now, for the given matrix $C=(c_{ij})\in M_n$, by the assumption,
there exist some $\{\delta_i,\gamma_i\}_{i=1}^n$ (depending on
$|x\rangle$) such that
$$C|x\rangle=\sum_{i=1}^n\delta_i(\sqrt{n-1}E_{ii})|x\rangle
+\sum_{i=1}^n\gamma_iE_{i,i+1}|x\rangle.$$ It follows that
$$\sum_{j=1}^nc_{ij}x_j=\sqrt{n-1}\delta_ix_i+\gamma_ix_{i+1}$$
for each $i$, which implies that
$$\gamma_i=\sum_{i\not=j}c_{ij}\frac{x_j}{x_{i+1}}
+(c_{ii}-\sqrt{n-1}\delta_i)\frac{x_i}{x_{i+1}},\quad
i=1,2,\cdots,n.\eqno(3.5)$$ Let $F=F_x=\left(\begin{array}{cccccccc}
 \alpha_1& \alpha_2& \cdots &\alpha_n&\beta_1& \beta_2& \cdots &\beta_n\\
\delta_1& \delta_2& \cdots &\delta_n&\gamma_1& \gamma_2& \cdots
&\gamma_n
\end{array}\right)$. Note that
$$FF^\dag=\left(\begin{array}{cc}
 \sum_{i=1}^n(|\alpha_i|^2+|\beta_i|^2)&\sum_{i=1}^n(\alpha_i\bar{\delta_i}+\beta_i\bar{\gamma_i})\\
\sum_{i=1}^n(\bar{\alpha_i}\delta_i+\bar{\beta_i}\gamma_i)&\sum_{i=1}^n(|\delta_i|^2+|\gamma_i|^2)
\end{array}\right),$$
and $\|F\|>1$ if and only if $\|FF^\dag\|>1 $. So, to prove that
$W_\Phi$ is optimal, we only need to check that $\|FF^\dag\|>1$ for
some suitable $|x\rangle$ and any choice of the coefficient matrix
$F=F_x$.

{\bf Case 1.}  ${\rm Tr}(C)\not=0$ or there exists at least one of
$c_{ij}$ with $i\not=j$ such that $c_{ij}\not=0$.

We consider the special case that $r_i=1$ for all $i$ in Eq.(3.4).
Then it follows from Eq.(3.4) that
$$\begin{array}{rl}
f(\alpha_1,\alpha_2,\cdots,\alpha_n)
\geq&n\sum_{i=1}^na_i^2+n-2\sqrt{n-1}\sum_{i=1}^na_i\\
=&\sum_{i=1}^n(na_i^2-2\sqrt{n-1}a_i+1).\end{array}$$ Let
$g(t)=nt^2-2\sqrt{n-1}t+1$. It is easily checked that $g$ attains
its minimum $\frac{1}{n}$ at the point $t_0=\frac{\sqrt{n-1}}{n}$.
So $f(\alpha_1,\alpha_2,\cdots,\alpha_n)\geq 1$ and attains its
minimum 1 at the point
$(\frac{\sqrt{n-1}}{n},\frac{\sqrt{n-1}}{n},\cdots,\frac{\sqrt{n-1}}{n})$.
Thus the best contractive coefficient matrix  is
$$\begin{array}{rl}&(\alpha_1,\alpha_2,\cdots,\alpha_n,\beta_1,\beta_2,\cdots,\beta_n)\\
=&(\frac{\sqrt{n-1}}{n},\frac{\sqrt{n-1}}{n},\cdots,\frac{\sqrt{n-1}}{n},
\frac{ {1}}{n}\frac{x_1}{x_{2}},
\frac{{1}}{n}\frac{x_2}{x_{3}},\cdots,\frac{{1}}{n}\frac{x_n}{x_{1}}),\end{array}\eqno(3.6)$$
and $\sum_{i=1}^n(|\alpha_i|^2+|\beta_i|^2)=1$. We may take
$x_i=e^{i\theta_i}$ for each $i$ as $r_i=1$ for each $i$. Thus
$|x\rangle=(e^{i\theta_1},e^{i\theta_2},\cdots,e^{i\theta_n})^T$,
and, for such $|x\rangle$, Eq.(3.5) becomes
$$\gamma_i=\sum_{i\not=j}c_{ij}e^{i(\theta_j-\theta_{i+1})}
+(c_{ii}-\sqrt{n-1}\delta_i)e^{i(\theta_i-\theta_{i+1})},\quad
i=1,2,\cdots,n.\eqno(3.7)$$ By Eqs.(3.6)-(3.7), one obtains
$$\sum_{i=1}^n(\alpha_i\overline{\delta_i}+\beta_i\overline{\gamma_i})
=\frac{1}{n}(\sum_{i=1}^nc_{ii}+\sum_{i\not=j}\overline{c_{ij}}e^{i(\theta_i-\theta_j)}).$$
As $\sum_{i=1}^nc_{ii}={\rm Tr}(C)\not=0$ or $c_{ij}\not=0$ for some
$i,j$ with $i\not=j$, we can choose suitable $\theta_i$s such that
$\sum_{i=1}^n(\alpha_i\overline{\delta_i}+\beta_i\overline{\gamma_i})\not=0$.
It follows, by Lemma 3.1,  that $\|FF^\dag\|>1$ for any choice of
the coefficient $(\delta_1,\ldots, \delta_n, \gamma_1,
\ldots,\gamma_n) $. Hence $\Psi_C$ is not positive by Lemma 2.1.

{\bf Case 2.}  ${\rm Tr}(C)=0$ and $c_{ij}=0$ for all $1\leq
i\not=j\leq n$.

In this case, $C={\rm diag}(c_{11},\ldots , c_{nn})$ with
$\sum_{i=1}^nc_{ii}=0$, and,
 Eq.(3.4) implies
$$\begin{array}{rl}
f(\alpha_1,\alpha_2,\cdots,\alpha_n)
=&\sum_{i=1}^na_i^2+\sum_{i=1}^nb_i^2+\sum_{i=1}^n(n-1)a_i^2r_i\\&
+\sum_{i=1}^n(n-1)b_i^2r_i+\sum_{i=1}^nr_i-2\sqrt{n-1}\sum_{i=1}^na_ir_i\\
\geq&\sum_{i=1}^n(1+(n-1)r_i)a_i^2
+\sum_{i=1}^nr_i-2\sqrt{n-1}\sum_{i=1}^na_ir_i\\
=&\sum_{i=1}^n((1+(n-1)r_i)a_i^2
+r_i-2\sqrt{n-1}a_ir_i).\end{array}$$ Notice that the function
$h(t)=(1+(n-1)r)t^2 -2\sqrt{n-1}rt+r$ achieves its minimum
$\frac{r}{1+(n-1)r}$ at $t_0=\frac{\sqrt{n-1}r}{1+(n-1)r}$. Hence
$f(\alpha_1,\alpha_2,\cdots,\alpha_n)$ attains its minimum
$\sum_{i=1}^n\frac{r_i}{1+(n-1)r_i}$ at the point
$(\alpha_1,\ldots,\alpha_n)=(\frac{\sqrt{n-1}r_1}{1+(n-1)r_1},\ldots,\frac{\sqrt{n-1}r_n}{1+(n-1)r_n})$.
Together with Eq.(3.3), we see that the coefficient matrix
$$\begin{array}{rl}&(\alpha_1,\alpha_2,\cdots,\alpha_n,\beta_1,\beta_2,\cdots,\beta_n)\\
=&(\frac{\sqrt{n-1}r_1}{1+(n-1)r_1},\frac{\sqrt{n-1}r_2}{1+(n-1)r_2},\cdots,\frac{\sqrt{n-1}r_n}{1+(n-1)r_n},\\&
\frac{\sqrt{r_1}}{1+(n-1)r_1}e^{i(\theta_1-\theta_2)},
\frac{\sqrt{r_2}}{1+(n-1)r_2}e^{i(\theta_2-\theta_3)},\cdots,
\frac{\sqrt{r_n}}{1+(n-1)r_n}e^{i(\theta_n-\theta_1)})\end{array}\eqno(3.8)$$
attains the minimal norm for
$|x\rangle=(|x_1|e^{i\theta_1},|x_2|e^{i\theta_2},\cdots,|x_n|e^{i\theta_n})^T$
with $r_i=|\frac{x_i}{x_{i+1}}|^2$, $i=1,2,\cdots,n$.

For the given $C={\rm diag}(c_{11}, \ldots ,c_{nn})$, write
$c_{ii}=s_i+it_i$. Let
$(\delta_1,\ldots,\delta_n,\gamma_1,\ldots,\gamma_n) $ be the
associated coefficients of $C$ at the above vector $|x\rangle$.
Write $\delta_j$ in the form $\delta_j=u_j+iv_j$. Consider the
function
$f_C(\delta_1,\delta_2,\cdots,\delta_n)=\sum_{i=1}^n(|\delta_i|^2+|\gamma_i|^2)$.
By Eq.(3.5), we have
$$\begin{array}{rl}
f_C(\delta_1,\delta_2,\cdots,\delta_n)
=&\sum_{i=1}^nu_i^2+\sum_{i=1}^nv_i^2+\sum_{i=1}^n(n-1)u_i^2r_i+\sum_{i=1}^ns_i^2r_i\\&
+\sum_{i=1}^n(n-1)(n-1)v_i^2r_i+\sum_{i=1}^nt_i^2r_i\\&-2\sqrt{n-1}\sum_{i=1}^nu_is_ir_i-2\sqrt{n-1}\sum_{i=1}^nv_it_ir_i\\
=&\sum_{i=1}^n [(1+(n-1)r_i)u_i^2-2\sqrt{n-1}u_is_ir_i+s_i^2r_i]\\&+
\sum_{i=1}^n
[(1+(n-1)r_i)v_i^2-2\sqrt{n-1}v_it_ir_i+t_i^2r_i].\end{array}\eqno(3.9)$$
Consider the function
$$h_C(y)=(1+(n-1)r)y^2-2\sqrt{n-1}sry+s^2r.$$
A simple calculation shows that $h_C$ attains the minimum
$\frac{rs^2}{1+(n-1)r}$ at the point
$y_0=\frac{\sqrt{n-1}rs}{1+(n-1)r}$. Thus, by Eq.(3.9), we get
$$f_C(\delta_1,\delta_2,\cdots,\delta_n)\geq
\sum_{i=1}^n\frac{\sqrt{n-1}r_i|c_i|^2}{1+(n-1)r_i}.$$ Moreover,
$f_C(\delta_1,\delta_2,\cdots,\delta_n)$ achieves its minimum at
$$(\delta_1,\delta_2,\cdots,\delta_n)
=(\frac{\sqrt{n-1}r_1c_1}{1+(n-1)r_1},\frac{\sqrt{n-1}r_2c_2}{1+(n-1)r_2},\cdots,\frac{\sqrt{n-1}r_nc_n}{1+(n-1)r_n}),$$
and the associated coefficient matrix is
$$\begin{array}{rl}&(\delta_1,\delta_2,\cdots,\delta_n,\gamma_1,\gamma_2,\cdots,\gamma_n)\\
=&(\frac{\sqrt{n-1}r_1c_1}{1+(n-1)r_1},\frac{\sqrt{n-1}r_2c_2}{1+(n-1)r_2},\cdots,\frac{\sqrt{n-1}r_nc_n}{1+(n-1)r_n},\\&
\frac{\sqrt{r_1}c_1}{1+(n-1)r_1}e^{i(\theta_1-\theta_2)},
\frac{\sqrt{r_2}c_2}{1+(n-1)r_2}e^{i(\theta_2-\theta_3)},\cdots,
\frac{\sqrt{r_n}c_n}{1+(n-1)r_n}e^{i(\theta_n-\theta_1)}).\end{array}\eqno(3.10)$$
By Eq.(3.8) and (3.10), we get
$$FF^\dag=\left(\begin{array}{cc}
\sum_{i=1}^n\frac{r_i}{1+(n-1)r_i}& \sum_{i=1}^n\frac{r_i\bar{c_i}}{1+(n-1)r_i}\\
 \sum_{i=1}^n\frac{r_ic_i}{1+(n-1)r_i}&\sum_{i=1}^n\frac{r_i|c_i|^2}{1+(n-1)r_i}\end{array}\right).$$
Let
$$|y_{(r_1,r_2,\cdots,r_n)}\rangle=FF^\dag
 \left(\begin{array}{c}
1\\
0\end{array}\right)
 = \begin{pmatrix}
\sum_{i=1}^n\frac{r_i}{1+(n-1)r_i}\\
 \sum_{i=1}^n\frac{r_ic_i}{1+(n-1)r_i}\end{pmatrix}.$$
Then
$$\begin{array}{rl}
&\||y_{(r_1,r_2,\cdots,r_n)}\rangle\|^2\\=
&(\sum_{i=1}^n\frac{r_i}{1+(n-1)r_i})^2+|\sum_{i=1}^n\frac{r_ic_i}{1+(n-1)r_i}|^2\\
=&(\sum_{i=1}^n\frac{r_i}{1+(n-1)r_i})^2+(\sum_{i=1}^n\frac{r_is_i}{1+(n-1)r_i})^2
+(\sum_{i=1}^n\frac{r_it_i}{1+(n-1)r_i})^2.\end{array}\eqno(3.11)$$
Since $\sum_{i=1}^nc_{ii}=0$, there exists at least a number $t_i$,
say $t_n$, such that $t_1+t_2+\cdots+t_{n-1}=-t_n>0$ (or there
exists at least  one $s_i$, say $s_n$, such that $s_1+\cdots
+s_{n-1}=-s_n>0)$. Assume, without loss of generality, that
$t_0=t_1+t_2+\cdots+t_{n-1}=-t_n>0$. Note that $\prod_{i=1}^nr_i=1$
with $r_i>0$ for each $i$. Let $r_i\rightarrow\infty$ for
$i=1,2,\ldots,n-1$, then we have $r_n\rightarrow 0$. Since
$$\lim_{r_i\rightarrow\infty}\frac{r_i}{1+(n-1)r_i}=\frac{1}{n-1}\
(i=1,2,\cdots, n-1)\quad{\rm and}
 \lim_{r_n\rightarrow 0}\frac{r_n}{1+(n-1)r_n}=0,$$
for any $\varepsilon>0$, there exists $N$ such that
$0<\frac{1}{n-1}-\frac{r_i}{1+(n-1)r_i}<\varepsilon$  for
$i=1,\ldots,n-1$ and $\frac{r_n}{1+(n-1)r_n}<\varepsilon$ whenever
$r_i\geq N$, $i=1,2,\ldots,n-1$. Thus we can write
$$\frac{r_i}{1+(n-1)r_i}=\frac{1}{n-1}-\delta_{r_i}\varepsilon
\quad{\rm with} \quad i=1,2,\cdots, n-1\eqno(3.12)$$and $$
\frac{r_n}{1+(n-1)r_n}=\delta_{r_n}\varepsilon\eqno(3.13)$$ with
$0<\delta_{r_i}<1$   for each $i$. Combining Eqs.(3.11)-(3.13), one
obtains
$$\begin{array}{rl}
&\||y_{(r_1,r_2,\cdots,r_n)}\rangle\|^2\\
\geq&[\sum_{i=1}^{n-1}(\frac{1}{n-1}-\delta_{r_i}\varepsilon)+\delta_{r_n}\varepsilon]^2
+[\sum_{i=1}^{n-1}(\frac{t_i}{n-1}-\delta_{r_i}t_i\varepsilon)+\delta_{r_n}t_n\varepsilon]^2\\
=&[1-(\sum_{i=1}^{n-1}\delta_{r_i}-\delta_{r_n})\varepsilon]^2
+[\frac{t_0}{n-1}-(\sum_{i=1}^{n-1}\delta_{r_i}t_i+\delta_{r_n}t_0)\varepsilon]^2\\
=&1+\mu^2\varepsilon^2-2\mu\varepsilon+\frac{t_0^2}{(n-1)^2}
+\nu^2\varepsilon^2-2\frac{t_0\nu}{n-1}\varepsilon\\
=& 1+
\frac{t_0^2}{(n-1)^2}+(\mu^2\varepsilon-2\mu+\nu^2\varepsilon
-2\frac{t_0\nu}{n-1})\varepsilon\\
\geq &  1+
\frac{t_0^2}{(n-1)^2}-2(\mu+\frac{t_0\nu}{n-1})\varepsilon,\end{array}\eqno(3.14)$$
where $\mu=\sum_{i=1}^{n-1}\delta_{r_i}-\delta_{r_n}$ and
$\nu=\sum_{i=1}^{n-1}\delta_{r_i}t_i+\delta_{r_n}t_0$. Note that
$|\mu|\leq n$ and $|\nu|\leq \sum_{i=1}^n
|t_i|\leq\sum_{i=1}^n|c_{ii}|$. Thus, if we take $\varepsilon$ so
that $$\varepsilon <
\frac{t_0^2}{2(n-1)(n(n-1)+t_0\sum_{i=1}^n|c_{ii}|)},$$ then, by
Eq.(3.14) we see that
$$\begin{array}{rl}\||y_{(r_1,r_2,\cdots,r_n)}\rangle\|^2
\geq & 1+
\frac{t_0^2}{(n-1)^2}-2(\mu+\frac{t_0\nu}{n-1})\varepsilon \\ \geq
& 1+
\frac{t_0^2}{(n-1)^2}-2(n+\frac{t_0\sum_{i=1}^n|c_{ii}|}{n-1})\varepsilon
>1. \end{array}$$ It follows that, for those $|x\rangle$ with $r_i\geq N$, $i=1,2,\ldots, n-1$,
 $\|FF^\dag\|>1$ for any possible choice of
the coefficients. Hence $\Psi_C$ is again not positive.

Now, by use of Corollary 2.3, $W_\Phi$ is optimal, completing the
proof of the assertion (1).

 Next let us prove the assertion (2). Let $n\geq 4$ be an even integer.
One can check that, the entanglement witness
$W_{\Phi^{(n,\frac{n}{2})}}$ has the form
$W_{\Phi^{(n,\frac{n}{2})}}=P+Q^{ T}$, where
$$P=\sum_{i=1}^n
(n-2)E_{ii}\otimes E_{ii} -\sum _{i\not= j, |i-j|\not=\frac{n}{2}}
E_{ij}\otimes E_{ij}\geq 0,
$$and
$$Q=\sum_{i=1}^n
E_{i+\frac{n}{2},i+\frac{n}{2}}\otimes E_{ii}-\sum
_{i=1}^{\frac{n}{2}}(E_{i+\frac{n}{2},i}\otimes
E_{i,i+\frac{n}{2}}+E_{i,i+\frac{n}{2}}\otimes
E_{i+\frac{n}{2},i})\geq 0.
$$
Hence  $W_{\Phi^{(n,\frac{n}{2})}}$ is decomposable, and not optimal
as $P\not= 0$. As a consequence, we see that the positive map
$\Phi^{(n,\frac{n}{2})}$ is decomposable.

The proof of Theorem 3.2 is finished. \hfill$\Box$

{\bf Remark 3.3.} We remark here that our $W_{\Phi^{(n,n-1)}}$
coincides with $W_{(n,1)}$ discussed in \cite{ABCL}, where the
authors asked whether $W_{(n,1)}$ is optimal. Theorem 3.2 gives an
affirmative answer to this problem. It is worth to noting that it
was shown in \cite{ABCL} that $W_{(n,1)}$ has no spanning property,
that is, span${\mathcal P}_{W_{(n,1)}}\not={\mathbb C}^n\otimes
{\mathbb C}^n$. In fact all $W_{\Phi^{(n,k)}}$s have no spanning
property. Assume that $|x\rangle\otimes
|y\rangle=(x_1y_1,x_2y_1,\ldots, x_ny_1,x_1y_2,\ldots, x_ny_n)^T$ is
a product vector from ${\mathbb C}^n\otimes {\mathbb C}^n$ such that
on which $W_{\Phi^{(n,k)}}$ has zero mean values, where
$|x\rangle=(x_1,\ldots, x_n)^T$ and $|y\rangle=(y_1,\ldots, y_n)^T$.
If all $y_i$ are nonzero, then it is not too difficult to check that
$|x\rangle\otimes |y\rangle$ has the form of
$$ \sum_{k=1}^n e^{i\theta_k}|k\rangle \otimes \sum_{k=1}^n
e^{-i\theta_k}|k\rangle.
$$
These states span a subspace
$$L=\{(\xi_{11},\xi_{21},\ldots, \xi_{n1},\xi_{12},\xi_{22}\ldots, \xi_{nn})^T :\xi_{ij}\in{\mathbb C}, \xi_{11}=\xi_{22}=\cdots=\xi_{nn}\}  $$
 of ${\mathbb
C}^n\otimes {\mathbb C}^n$. If $y_i=0$ for some $i$, then one may
check that $y_j\not=0$ implies that $x_j=0$. This forces that
$\xi_{ii}=x_iy_i=0$ and hence $|x\rangle\otimes |y\rangle\in L$.
Therefore, $L$ is in fact the subspace spanned by all product
vectors on which $W_{\Phi^{(n,k)}}$ has zero mean values. As $\dim
L=n^2-n+1<n^2$, $W_{\Phi^{(n,k)}}$ has no spanning property.

\if {\bf Proposition 3.4.} {\it Let $a_1, a_2,\ldots, a_n$ be
nonnegative numbers with $\sum_{i=1}^na_i=n-1$, $a_1\geq 1$ and at
least one of the rest  nonzero. Let $\Phi:M_n\rightarrow M_n$ be a
linear map defined by
$$\begin{array}{rl} \Phi(X)=&{\rm diag}((1+a_1)x_{11}+a_2x_{22}+\cdots +a_nx_{nn}, a_nx_{11}+(1+a_1)x_{22}+\cdots
+a_{n-1}x_{nn},\\
&\cdots, a_2x_{11}+a_3x_{22}+\cdots +(1+a_1)x_{nn})-X\end{array}$$
for any $X=(x_{ij})\in M_n$. Then $W_\Phi$ has no spanning
property.}\fi

\section{Extending the set of witnesses which support the SPA conjecture}

Now let us turn to another topic so-called structural physical
approximation (SPA) \cite{ABCL,KABL}. Among the criteria of
separability of states,  probably the most powerful one is the
positive map criterion: a given state $\rho$ acting on $M_n({\mathbb
C})\otimes M_m({\mathbb C})$ is separable if and only if, for any
positive map $\Phi:M_n\rightarrow M_m$, the operator
$(I_n\otimes\Phi )\rho$ is positive. Despite its proven efficiency
in entanglement detection, the positive map criterion of
separability above is not directly applicable in experiments as the
NCP positive maps do not represent physical processes. So it is
important to design methods which could make the experimental
detection of entanglement with the aid of positive maps possible.
SPA is one of such methods. Let $D:M_n\rightarrow M_m$ be the
completely depolarizing channel, i.e., $D(\rho)={\rm
Tr}(\rho)I_m/m$. $D$ is certainly an interior of the convex set of
positive maps from $M_n$ into $M_m$. It is clear $D$ is completely
positive. For any positive map $\Phi:M_n\rightarrow M_m$, let
$$\tilde{\Phi}[p]=(1-p)D+p\Phi \quad( 0\leq
p\leq 1).\eqno(4.1)$$ From the C-J isomorphism it is clear that
there exists a $p_*\in(0,1)$ such that $\Phi[p]$ is completely
positive whenever $0\leq p\leq p_*$ (if $\Phi$ is trace-preserving,
$\tilde{\Phi}[p]$ represents a quantum channel whenever $0\leq p\leq
p_*$), and thus, in principle, it can represent some physical
process. The least noisy completely positive map from the class
$\tilde{\Phi}[p]$ ($0\leq p\leq p_*$), i.e., $\tilde{\Phi}[p_*]$ is
called the structural physical approximation of $\Phi$.

Recall that a completely positive map $\Phi$ is said to be
entanglement breaking if $I\otimes \Phi$ sends all states to
separable states.

The following conjecture is posed in \cite{KABL}. Here, $\Phi$ is
said to be  optimal if the corresponding entanglement witness
$W_\Phi$ is optimal.

{\bf Conjecture 4.1.} {\it Let $\Phi$ be an optimal
(trace-preserving) positive map. Then its SPA is entanglement
breaking map (channel).}

Applying C-J isomorphism, one can define SPA of an entanglement
witness: Let $W$ be a normalized EW, i.e., ${\rm Tr}(W)=1$. An
operator $\tilde{W}(p)$ defined by
$$\tilde{W}(p)=(1-p)\frac{I_n\otimes I_m}{mn}+pW \quad (0\leq
p\leq 1)
$$
is called structural physical approximation (SPA) of $W$ if
$\tilde{W}(p)\geq 0$. The maximal value  of such $p$ is given by
$p_*=1/(1+mn\lambda)$, where $-\lambda<0$ is the smallest eigenvalue
of $W$. Thus Conjecture 4.1 reformulated as

{\bf Conjecture 4.1$^\prime$.} {\it Let $W$ be an optimal
entanglement witness with ${\rm Tr}(W)=1$. Then $\tilde{W}(p_*)$
defines a separable state.}

Conjecture 4.1 (4.1$^\prime$) is supported by several examples (ref.
\cite{ABCL,KABL}. We shall show that the normalized optimal
indecomposable entanglement witnesses
$W^{(n,k)}=\frac{1}{n(n-1)}W_{\Phi^{(n,k)}}$ ($k\not= \frac{n}{2}$)
support the above conjecture,too.

{\bf Proposition 4.2.} {\it Let $n\geq 3$ and $1\leq k\leq n-1$ with
$k\not=\frac{n}{2}$. Then $\tilde{W}^{(n,k)}(p_*)$ is a separable
state.}

{\bf Proof.} It is clear that the minimal eigenvalue of $W^{(n,k)}$
is $-\frac{1}{n(n-1)}$. So $p_*=\frac{n-1}{2n-1}$. Thus
$$\begin{array}{rl}\tilde{W}^{(n,k)}(p_*)=&(1-p_*)\frac{I_n\otimes I_n}{n^2}+p_*{W}^{(n,k)} \\
=&\frac{1}{n(2n-1)}(I_{n^2}+\sum_{i=1}^n(n-2)E_{ii}\otimes
E_{ii}+\sum_{i=1}^nE_{i+(n-k),i+(n-k)}\otimes
E_{ii}\\&-\sum_{i=1}^nE_{i,i+k}\otimes
E_{i,i+k}-\sum_{i=1}^nE_{i+k,i}\otimes E_{i+k,i})\\
=&\frac{1}{n(2n-1)}(\sum_{i\not=j}E_{ii}\otimes
E_{jj}+\sum_{i=1}^nE_{i+(n-k),i+(n-k)}\otimes
E_{ii}\\&+(n-1)\sum_{i=1}^nE_{ii}\otimes
E_{ii}-\sum_{i=1}^nE_{i,i+k}\otimes
E_{i,i+k}-\sum_{i=1}^nE_{i+k,i}\otimes E_{i+k,i}).\end{array}$$
Since $\sum_{i\not=j,|i-j|\not=k}E_{ii}\otimes
E_{jj}+\sum_{i=1}^nE_{i+(n-k),i+(n-k)}\otimes E_{ii}$ is
non-normalized separable density matrix, we only need to prove that
$$\sigma=(n-1)\sum_{i=1}^nE_{ii}\otimes E_{ii}-\sum_{i=1}^nE_{i,i+k}\otimes
E_{i,i+k}-\sum_{i=1}^nE_{i+k,i}\otimes E_{i+k,i}$$ is separable.
However, $\sigma=\sum_{i=1}^n\sigma_{i,i+k}$ with
$\sigma_{i,i+k}=E_{ii}\otimes E_{ii}+E_{i+k}\otimes
E_{i+k}+E_{i,i}\otimes E_{i+k,i+k}+E_{i+k,i+k}\otimes
E_{ii}-E_{i,i+k}\otimes E_{i,i+k}-E_{i+k,i}\otimes E_{i+k,i}$. Note
that $\sigma_{i,i+k}$ stands for two-qubit matrix embedded in
$M_{n^2}$ and has positive partial transposition. So
$\sigma_{i,i+k}$ is separable, which implies that $\sigma$ is also
separable. Therefore, $\tilde{W}^{(n,k)}(p_*)$ is a separable state,
as desired. \hfill$\Box$

\section{Optimal entanglement witnesses for infinite dimensional systems}

Based on the results in Section 3, we can obtain some optimal
indecomposable entanglement witnesses for infinite dimensional
systems.

 Let $H$ and $K$ be separable infinite dimensional Hilbert
spaces, and, for any positive integer $n\geq 3$, let
$\{|i\rangle\}_{i=1}^n$ and $\{|j'\rangle\}_{j=1}^n$ be any
orthonormal sets of $H$ and $K$, respectively. For each
$k=1,2,\cdots, n-1$, let $\hat{\Phi}^{(n,k)}:{\mathcal
B}(H)\rightarrow {\mathcal B}(K)$ be defined by
$$\begin{array}{rl} \hat{\Phi}^{(n,k)}
(A)=&(n-1)\sum_{i=1}^nE_{ii}AE_{ii}^\dagger+\sum_{i=1}^{n}E_{i,\pi^k(i)}AE_{i,\pi^k(i)}^\dagger\\&-
(\sum_{i=1}^nE_{ii})A(\sum_{i=1}^nE_{ii})^\dagger
\end{array}\eqno(5.1)$$
for every $A\in{\mathcal B}(H)$, where $\pi(i)=\pi^1(i)=(i+1)\ {\rm
mod} \ n$, $\pi^k(i)=(i+k)\ {\rm mod} \ n$ ($k>1$), $i=1,2,\cdots,
n$ and $E_{ji}=|j'\rangle\langle i|$. It is shown in \cite{QH} that,
$\hat{\Phi}^{(n,k)}$s  are  NCP  positive linear maps. Moreover,
$\hat{\Phi}^{(n,k)}$ is indecomposable whenever $k\not=\frac{n}{2}$.

Let $P_n^+=|\psi_n\rangle\langle\psi_n|$, where
$|\psi_n\rangle=|11\rangle+|22\rangle+\cdots +|nn\rangle$, and let
$$\hat{W}_{\hat{\Phi}^{(n,k)}}=(I\otimes\hat{\Phi}^{(n,k)}
)P^+_n.\eqno(5.2)$$ Then $\hat{W}_{\hat{\Phi}^{(n,k)}}$ is an
entanglement witness for the system living in $H\otimes K$.

{\bf Theorem 5.1.} {\it Let $H$ and $K$ be infinite dimensional
complex Hilbert spaces. For any positive integers $n\geq 3$ and $k$,
let $\hat{\Phi}^{(n,k)}$ and $\hat{W}_{\hat{\Phi}^{(n,k)}}$ be the
positive  maps and the entanglement witnesses defined in Eq.(5.1)
and Eq.(5.2), respectively.}

(1) {\it $\hat{W}_{\hat{\Phi}^{(n,k)}}$ is  indecomposable and
optimal whenever $k\not=\frac{n}{2}$.}

(2) {\it $\hat{W}_{\hat{\Phi}^{(n,\frac{n}{2})}}$ is decomposable
and not optimal.}

{\bf Proof.} (1) Assume that $k\not=\frac{n}{2}$. Denote by $P$ and
$Q$ the $n$-rank projection with range the subspace spanned by
$\{|i\rangle\}_{i=1}^n$ and $\{|j'\rangle\}_{j=1}^n$. Then we have
$\hat{\Phi}^{(n,k)}(A)=Q[\hat{\Phi}^{(n,k)}(PAP)]Q $ holds for all
$A\in{\mathcal B}(H)$.

By \cite{HG}, an entanglement witness $W$ for an infinite
dimensional system is optimal if and only if $W-D$ can not be an
entanglement witness anymore for any nonzero operator $D\geq 0$. So,
if $\hat{W}_{\hat{\Phi}^{(n,k)}}$ is not optimal, then there exists
a nonzero positive operator $\hat{D}\in{\mathcal B}(H\otimes K)$
such that $\hat{W}_{\hat{\Phi}^{(n,k)}}-\hat{D}$ is an entanglement
witness. Note that $(P\times Q)\hat{W}_{\hat{\Phi}^{(n,k)}}(P\otimes
Q)=\hat{W}_{\hat{\Phi}^{(n,k)}}$. Then
$\hat{W}_{\hat{\Phi}^{(n,k)}}-\hat{D}$ is an entanglement witness
implies that, for any separable pure state $\sigma
_{lh}=|l\rangle\langle l|\otimes |h'\rangle\langle h'|$ with
$l,h>n$, we have
$$-{\rm Tr}(\hat{D}\sigma_{lh})={\rm
Tr}((\hat{W}_{\hat{\Phi}^{(n,k)}}-\hat{D})\sigma_{lh})\geq 0,
$$
which forces that
$$ \langle lh'|\hat{D}|lh'\rangle=0$$
for all $l,h>n$. Since $\hat{D}\geq 0$, we see that
$$ \hat{D}=(P\otimes Q)\hat{D}(P\otimes Q).$$

Observe that $\hat{W}_{\hat{\Phi}^{(n,k)}}|_{(P\otimes Q)(H\otimes
K)}= W_{\Phi^{(n,k)}}$ with $W_{\Phi^{(n,k)}}$ the same as that in
Theorem 3.2. Denote by $D=\hat{D}|_{(P\otimes Q)(H\otimes K)}$. Then
$\hat{D}=D\oplus 0$. Now, for any separable state
$\sigma\in{\mathcal S}(P(H)\otimes Q(K))$, there exists a separable
state $\sigma'\in{\mathcal S}(H\otimes K)$ with $\sigma'=(P\otimes
Q)\sigma'(P\otimes Q)$ such that $\sigma=\sigma'|_{(P\otimes
Q)(H\otimes K)}$. Then we have
$$ {\rm Tr}((W_{\Phi^{(n,k)}}-D)\sigma)={\rm
Tr}((\hat{W}_{\hat{\Phi}^{(n,k)}}-\hat{D})\sigma')\geq 0,
$$
which means that $W_{\Phi^{(n,k)}}-D$ is an entanglement witness,
contradicting to the fact that $W_{\Phi^{(n,k)}}$ is optimal. Hence
$\hat{W}_{\hat{\Phi}^{(n,k)}}$ is optimal, completing the proof of
the statement (1).

The proof statement (2) is the same as that of (2) in Theorem 3.2.
\hfill$\Box$

{\bf Remark 5.2.} By checking the proof of Theorem 5.1, one sees
that the following general result is true: Let $W$ be an
entanglement witness on $H\otimes K$. If there exist projections
$P\in{\mathcal B}(H)$ and $Q\in{\mathcal B}(K)$ such that
$W=(P\otimes Q)W(P\otimes Q)$ and $W|_{(P\otimes Q)(H\otimes K)}$ is
optimal on $P(H)\otimes Q(K)$, then $W$ is optimal.

It was asked in \cite{QH} whether or not
$\hat{\Phi}^{(n,\frac{n}{2})}$ is indecomposable. The following
proposition gives an answer to this question.

{\bf Proposition 5.3.} {\it Let $H$ and $K$ be infinite dimensional
complex Hilbert spaces. For any even positive integers $n\geq 4$,
let $\hat{\Phi}^{(n,\frac{n}{2})}$ be the positive  map defined in
Eq.(5.1). Then $\hat{\Phi}^{(n,\frac{n}{2})}$ is decomposable.}

{\bf Proof.} Note that, for infinite dimensional systems,  we have
no one-to-one correspondence between the set of positive linear maps
and the set of entanglement witnesses complemented by
Choi-Jamio{\l}kowski isomorphism. So, we can not get the
decomposability of $\hat{\Phi}^{(n,\frac{n}{2})}$ from the
decomposability of $\hat{W}_{\hat{\Phi}^{(n,\frac{n}{2})}}$ proved
in Theorem 5.1.

Let $P$ and $Q$ be as that in the proof of Theorem 5.1. Then
$$\hat{\Phi}^{(n,\frac{n}{2})}
(A)=Q\hat{\Phi}^{(n,\frac{n}{2})}(PAP)Q$$ for all $A\in{\mathcal
B}(H)$. Thus, restricting to ${\mathcal B}(P(H)\otimes Q(K))$ we get
$\hat{\Phi}^{(n,\frac{n}{2})}|_{{\mathcal B}(P(H)\otimes
Q(K))}=\Phi^{(n,\frac{n}{2})}$. Since  the positive map $\Phi:
{\mathcal B}(P(H))\rightarrow {\mathcal B}(Q(K))$ and its associated
entanglement witness $W_\Phi$ is one-to-one corresponded, and $\Phi$
is decomposable if and only if $W_\Phi$ is decomposable. By applying
Theorem 3.2, $W_{\Phi^{(n,\frac{n}{2})}}$ is decomposable. So
$\Phi^{(n,\frac{n}{2})}$ is decomposable. Thus, there are completely
positive maps $\Delta_i$ ($i=1,2$) such that
$\Phi^{(n,\frac{n}{2})}(A)=\Delta_1 (A)+\Delta_2(A^T)$ for all
$A\in{\mathcal B}(P(H))$, where the transpose is taken with respect
to the basis $\{|i\rangle\}_{i=1}^n$. Extending $\Delta_i$ to
$\hat{\Delta}_i$ on ${\mathcal B}(H)$ by
$\hat{\Delta}_i(A)=Q\Delta_i(PAP|_{P(H)})Q$ regarding $Q$ as an
operator from $K$ onto $Q(K)$. Then $\hat{\Delta}_i:{\mathcal
B}(H)\rightarrow{\mathcal B}(K)$ is completely positive for $i=1,2$
and $\hat{\Phi}^{(n,\frac{n}{2})}
(A)=\hat{\Delta}_1(A)+\hat{\Delta}_2(A^T)$, where the transpose is
taken with respect to the  given basis
$\{|j'\rangle\}_{j=1}^\infty$. Hence, $\hat{\Phi}^{(n,\frac{n}{2})}$
is decomposable.\hfill$\Box$

Furthermore, for each $i=1,2$, as $\Delta_i$ is an elementary
operator, we see that $\hat{\Delta}_i$ is an elementary operator,
too. So, there exist operators $C_1,\ldots , C_s, D_1, \ldots,
D_t\in {\mathcal B}(H,K)$ with $C_h=QC_hP$ and $D_l=QD_lP$ such that
$$\hat{\Phi}^{(n,\frac{n}{2})} (A)=\sum_{h=1}^s C_hAC_h^\dag
+(\sum_{l=1}^t D_lA^TD_l^\dag)
$$
for all $A$. By checking the proof of Proposition 5.3, we see that
this fact indeed hold for all finite rank positive elementary
operators. Thus a question is raised:

{\bf Question 5.4.} Let $\Phi :{\mathcal B}(H)\rightarrow{\mathcal
B}(K)$ be a decomposable positive elementary operator with infinite
rank. Whether or not there exist completely positive {\it
elementary} operators $\Delta_i$, $i=1,2$, such that $\Phi=\Delta_1+
\Delta_2\circ{\bf T}$?

\section{Conclusions}

We present a characterization of optimal entanglement witnesses in
terms of positive maps and provide a general approach of checking
optimality of entanglement witnesses. This allows us to show that a
kind of indecomposable entangled witnesses $W_{\Phi^{(n,k)}}$
corresponding to the positive maps $\Phi^{(n,k)} : M_n\rightarrow
M_n$ with $k\not=\frac{n}{2}$ are optimal, where, for each $n\geq 3$
and $1\leq k\leq n-1$, $\Phi^{(n,k)}$ is defined by
$$\Phi^{(n,k)}(a_{ij})={\rm diag}((n-1)a_{11}+a_{1+k,1+k},
(n-1)a_{22}+a_{2+k,2+k},\cdots, (n-1)a_{nn}+a_{kk})-(a_{ij}).$$ The
space spanned by all product vectors on which $W_{\Phi^{(n,k)}}$ has
zero mean values is of dimension $n^2-n+1$. So, in addition to the
well-known indecomposable entanglement witness acting on $M_3\otimes
M_3$ corresponding to the Choi map, we get much more indecomposable
optimal EWs acting on $M_n\otimes M_n$ that have no spanning
property. These also allow us to get new examples of indecomposable
optimal EWs of infinite dimensional systems. Moreover, these optimal
EWs give new examples supporting a recent SPA conjecture posed in
\cite{KABL} saying that the so-called structural physical
approximations (SPA) to optimal positive maps (optimal EWs) give
entanglement breaking (EB) maps (separable states).


\end{document}